\documentclass[twocolumn,aps,floatfix,showpacs,nobibnotes,showkeys,amssymb,superscriptaddress]{revtex4}
\usepackage{graphicx,epsf,dcolumn,subfigure}
\usepackage{placeins}
\begin{document}

\def\baselinestretch{1.0}

\begin{titlepage}

\date{\today}

\title{Electronic structure and band parameters for Zn$X$ ($X$=O, S, Se, Te)}

\author{S. Zh. Karazhanov}

\affiliation{Department of Chemistry, University of Oslo, PO Box
1033 Blindern, N-0315 Oslo, Norway}

\affiliation{Physical-Technical Institute, 2B Mavlyanov St.,
700084 Tashkent, Uzbekistan}

\author{P. Ravindran}

\email[Corresponding author: Fax: +47 22 85 54 41/55 65,
]{E-mail:ponniah.ravindran@kjemi.uio.no}

\affiliation{Department of Chemistry, University of Oslo, PO Box
1033 Blindern, N-0315 Oslo, Norway}

\author{U. Grossner}

\affiliation{Department of Physics, University of Oslo, PO Box
1048 Blindern, N-0316 Oslo, Norway}

\author{A. Kjekhus}

\affiliation{Department of Chemistry, University of Oslo, PO Box
1033 Blindern, N-0315 Oslo, Norway}

\author{H. Fjellv{\aa}g}

\affiliation{Department of Chemistry, University of Oslo, PO Box
1033 Blindern, N-0315 Oslo, Norway}

\author{B. G. Svensson}

\affiliation{Department of Physics, University of Oslo, PO Box
1048 Blindern, N-0316 Oslo, Norway}

\begin{abstract}

First-principles density-functional calculations have been
performed for zinc monochalcogenides with zinc-blende- and
wurtzite-type structures. It is shown that the local-density
approximation underestimates the band gap, misplaces the energy
levels of the Zn-3$d$ states, and overestimates the crystal-field
splitting energy. Without spin-orbit coupling, the order of the
states at the top of VB is found to be normal for all the Zn$X$
phases considered. Upon inclusion of the spin-orbit coupling in
calculations, ZnO in zinc-blende- and wurtzite-type phases become
anomalous. It is shown that the Zn-3$d$ electrons are responsible
for the anomalous order. The effective masses of electrons and
holes have been calculated and found that holes are much
anisotropic and heavier than the electrons in agreement with
experimental findings. The typical errors in calculated band gaps
and related parameters originate from strong Coulomb correlations,
which are found to be highly significant in ZnO. The LDA+$U$
approach is found to correct the strong correlation of the Zn-3$d$
electrons, and thus improves the agreement with the experimentally
established location of the Zn-3$d$ levels. Consequently, it
increases significantly the parameters underestimated in the pure
LDA calculations.

\end{abstract}



\maketitle

\end{titlepage}

\def\baselinestretch{1.0}

\normalsize \vspace{-0.4cm}
\section{\label{intro} Introduction}
\vspace{-0.35cm}

Investigation of properties of zinc monochalcogenides [Zn$X$
($X$=O, S, Se, Te) with zinc-blende-(z-) and wurtzite-(w-)type
structure] has promoted much interest because of their numerous
applications in optoelectronic devices such as visual displays,
high-density optical memories, transparent conductors, solid-state
laser devices, photodetectors, solar cells etc. Despite frequent
studies over more than four decades many fundamental questions
still remain open. One of them is the eigenvalue problem related
to a severe underestimation of the band gaps ($E_g$), energy
levels of the Zn-3$d$ electrons ($E_d$) (for ZnO
[\onlinecite{UHKS02}], ZnS [\onlinecite{LICL02}], and ZnSe
[\onlinecite{LICL02}]), splitting energies of the states at the
top of the valence band (VB), spin-orbit (SO) coupling and
crystal-field (CF) splitting energies ($\Delta_{\rm{SO}}$ and
$\Delta_{\rm{CF}}$, respectively) calculated according to the
density-functional theory (DFT) within the local-density
approximation (LDA). These problems have been the subject of
numerous studies with different methods such as the LDA plus
self-interaction correction (LDA+$SIC$)
[\onlinecite{LRLSM02,VKP96,DPVASMACG04}], LDA plus the
multiorbital mean-field Hubbard potential (LDA+$U$)
[\onlinecite{F04}] (which includes the on-site Coulomb interaction
in the LDA Hamiltonian), and different version of GW approximation
[\onlinecite{UHKS02,LICL02,RQNFS05}]. In the latter approximation
"G" stands for one-particle Green's function as derived from
many-body perturbation theory and "W" for Coulomb screened
interactions. This approach takes into account both non-locality
and energy-dependent features of correlations in many-body system.
However, none of the above approaches has been able to remedy the
eigenvalue problem except the combination of exact-exchange DFT
calculations in the optimized-effective-potential approach with GW
[\onlinecite{RQNFS05}], which is found to give better agreement
with the experimental band gaps and the location of the Zn-3$d$
energy levels.

The order of the states at the top of the VB in ZnO-w is one of
the topics which is still frequently debated
[\onlinecite{RLJLCH99,LRLSM02}]. Another debated aspect is the
effective masses of the charge carriers although these are more
indefinite parameters for the Zn$X$ phases. At present the
effective masses from different \textit{ab initio} packages and
experiments scatter appreciably in publications on ZnO
[\onlinecite{LRLSM02,XC93,LYV96a,H73,IOTTK01}] and ZnTe
[\onlinecite{Madelung}]. In this work Zn$X$ ($X$=O, S, Se, Te) in
z- and w-type structural arrangements is studied by
first-principles calculations within the LDA, generalized gradient
approximation (GGA), and LDA+$U$ approaches. See
Ref.~[\onlinecite{KRGKFS051}] for a more detailed account of our
findings.

\vspace{-0.35cm}
\section{Computational details}
\vspace{-0.35cm}

The electronic band structure of the Zn$X$ phases is studied using
the VASP--PAW package [\onlinecite{vasp}], which calculates the
Kohn--Sham eigenvalues by the DFT~[\onlinecite{HK64}] within the
LDA~[\onlinecite{KS65}], GGA~[\onlinecite{PBE96}], and
LDA+$U$~[\onlinecite{ASKCS93,DBSHS98}] approximations. The
exchange and correlation energy per electron have been described
by the Perdew--Zunger parametrization~[\onlinecite{PZ81}] of the
quantum Monte Carlo procedure of
Ceperley--Alder~[\onlinecite{CA80}]. The interaction between
electrons and atomic cores is described by means of
non-norm-conserving pseudopotentials implemented in the VASP
package, which are generated in accordance to the
projector-augmented-wave (PAW) method [\onlinecite{PAW94,PAW99}].
Self-consistent calculations were performed using a $10\times
10\times 10$~mesh frame according to Monkhorst--Pack scheme for
z-type phases and to $\Gamma$-centered grids for w-type phases.
The completely filled semicore-Zn-3$d$ states have been considered
as valence states. Values for $\Delta_{\rm CF}$, $\Delta_{\rm
SO}$, and the average band gap $E_0$ are calculated within the
quasi-cubic model [\onlinecite{H60,MRR95}]. For band-structure
calculations we used the experimentally determined
crystal-structure parameters for all Zn$X$ phases considered.

Without SO coupling the top of the VB for phases with w-type
structure is split into a doublet $\Gamma_5$ and a singlet
$\Gamma_1$ state by the crystal-field. Inclusion of SO coupling
gives rise to three twofold degenerate valence bands, which are
denoted as $A$, $B$, and $C$ states. The symmetry of two of these
three bands are of $\Gamma_7$ character and one of $\Gamma_9$
character. Band gaps $E_g, E_A, E_B$, and $E_C$ are defined as the
difference between the conduction band (CB) minimum and energy
levels of the $A$, $B$, and $C$ states, respectively. Without SO
coupling the VB spectrum near the $\Gamma$~point for the Zn$X$-z
phases originates from the sixfold degenerate $\Gamma_{15}$ state.
The SO interaction splits the $\Gamma_{15}$ level into a fourfold
degenerate $\Gamma_8$ and doubly degenerate $\Gamma_7$ levels.

\vspace{-0.35cm}
\section{\label{Results} Results and discussion}
\subsection{\label{LDA+U} Value of the parameters $U$ and $J$}
\vspace{-0.35cm}

The simplified rotationally invariant LDA$+U$ approach
[\onlinecite{ASKCS93}] used in this work requires knowledge of the
values of the parameters $U$ and $J$. Since these parameters do
not explicitly take into account the final state, values of $U$
and $J$ were found empirically from LDA+$U$ band structure
calculations as a function of $U$ and $J$ such that the value of
$E_d$ obtained for particular $U$ and $J$ fit with the
experimentally determined location. For comparison, the values of
$U$ and $J$ have been calculated for some of the compounds within
the constrain DFT [\onlinecite{PEE98}], showing that the
calculated values to some extent agrees with those extracted
semiempirically.

\vspace{-0.4cm}
\subsection{\label{eigenvalue} Eigenvalues}
\vspace{-0.35cm}

\begin{figure}
\centering
\includegraphics[scale=0.6]{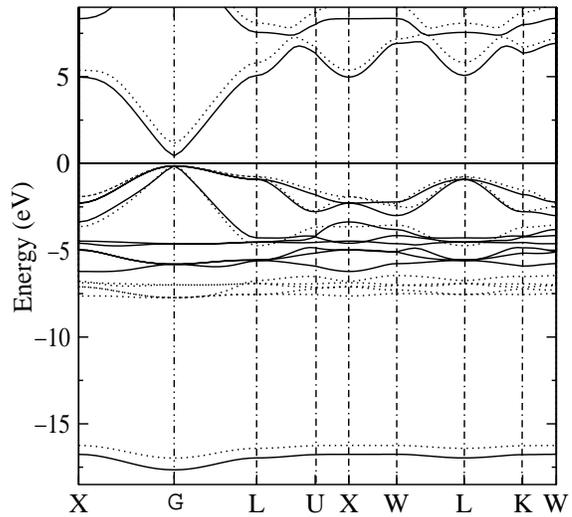}
\caption{Band dispersion for ZnO-z calculated within LDA (solid
lines) and LDA+$U$ (dotted lines) approaches. The Fermi level is
set at zero of energy.} \label{zZnO+disp}
\end{figure}

The results of the band structure calculations are listed in
Table~\ref{band-gap}. It is found that the band gaps ($E_g, E_A,
E_B, E_C$) and the mean energy level $E_d$ severely underestimated
in the LDA calculations. The error in the LDA calculated energies
is quite pronounced for ZnO (see e.g. Fig.~\ref{zZnO+disp} for
ZnO-z) compared to the other Zn$X$ phases and the discrepancy
exceeds the usual error for LDA calculations. Furthermore,
compared to experimental data the CF splitting energy
($\Delta_{\rm CF}$) is severely overestimated for ZnO-w by 2.4
times and underestimated for ZnS-w by around 1.2 times.
$\Delta_{\rm SO}$ is overestimated for ZnO-w to about 12.2 times,
underestimated for ZnS-w by about 3.2 times, but agrees well with
experimental data for ZnSe-z and ZnTe-z. Furthermore, $\Delta_{\rm
SO}$ is found to be negative for ZnO-z and -w phases in agreement
with results of
Refs.~\onlinecite{LRLSM02,LYV96a},~and~\onlinecite{RCP68,T60,SMK65},
while it is positive for the other Zn$X$ considered.

%%%%%%%%%%%%%%%%%%%%%%%%% Table 1  %%%%%%%%%%%%%%%%%%%%%%%%%%%%%%%%%%
\squeezetable
\begin{table*}
\caption{Band gaps $E_g, E_g(A), E_g(B), E_g(C),$ and $E_0$,
crystal-field ($\Delta_{\rm CF}^0, \Delta_{\rm CF}$), and
spin-orbit ($\Delta_{\rm SO}$) splitting energies (all in eV) for
Zn$X$ phases with z- and w-type structure calculated within LDA,
GGA, and LDA+$U$ approaches. $E_g$ and $\Delta_{\rm CF}^0$ refer
to calculations whithout the SO coupling, in all other
calculations the SO interactions are accounted for. Experimental
values are quoted when available.}
\begin{ruledtabular}
\begin{tabular}{llddddddddd}
\multicolumn{1}{l}{Phase} & \multicolumn{1}{l}{Method} & \multicolumn{1}{c}{$E_g$} & \multicolumn{1}{c}{$E_g(A)$} & \multicolumn{1}{c}{$E_g(B)$} & \multicolumn{1}{c}{$E_g(C)$} & \multicolumn{1}{c}{$E_0$} & \multicolumn{1}{c}{$E_d$} & \multicolumn{1}{c}{$\Delta_{\rm{CF}}^0$} & \multicolumn{1}{c}{$\Delta_{\rm{CF}}$} & \multicolumn{1}{c}{$\Delta_{\rm{SO}}$} \\
\hline
ZnO-w   & \rm{LDA}         & 0.7442 & 0.7238 & 0.7561 & 0.8385 & 0.773 & \sim 5.00  & 0.0951 & 0.0928 &-0.0429 \\
        & \rm{GGA}         & 0.8044 & 0.7831 & 0.8165 & 0.9000 & 0.833 & \sim 5.00  & 0.0967 & 0.0944 &-0.0443 \\
        & \rm{LDA+$U$}     & 1.9884 & 2.0080 & 2.0528 & 2.0530 & 2.038 & \sim 10.00 &        &        &        \\
& Expt.~[\onlinecite{MRR95}] &        & 3.4410 & 3.4434 & 3.4817 & 3.455 &            &        & 0.0394 &-0.0035 \\
ZnS-w   & \rm{LDA}         & 1.9896 & 1.9681 & 1.9947 & 2.0734 & 2.012 & \sim 6.50  & 0.0688 & 0.0518 & 0.0269 \\
        & \rm{GGA}         & 2.2322 & 2.2114 & 2.2361 & 2.3104 & 2.253 & \sim 6.00  & 0.0661 & 0.0494 & 0.0249 \\
        & \rm{LDA+$U$}     & 2.2828 & 2.2595 & 2.2858 & 2.3662 & 2.304 & \sim 8.20  & 0.0588 & 0.0549 & 0.0256 \\
&Expt.~[\onlinecite{LB2}]    &        & 3.8643 & 3.8932 & 3.9808 &       &            &        & 0.0580 & 0.0860 \\
&Expt.~[\onlinecite{LB2}]    &        & 3.8715 & 3.8998 &        &       &            &        & 0.0055 & 0.0920 \\
ZnSe-w  & \rm{LDA}         & 1.0704 & 0.9389 & 1.0080 & 1.3789 & 1.109 & \sim 6.50  & 0.1138 & 0.3243 & 0.0467 \\
        & \rm{GGA}         & 1.3271 & 1.2004 & 1.2678 & 1.6243 & 1.364 & \sim 6.50  & 0.1122 & 0.3105 & 0.0460 \\
        & \rm{LDA+$U$}     & 1.4039 & 1.2708 & 1.3336 & 1.7209 & 1.442 & \sim 9.30  & 0.1010 & 0.3465 & 0.0408 \\
&Expt.~[\onlinecite{LB2,Madelung}]&   & 2.8600 & 2.8760 & 2.9260 &       & \sim 9.20  &        &        &        \\
ZnTe-w  & \rm{LDA}         & 1.0519 & 0.7600 & 0.8200 & 1.6912 & 1.091 & \sim 7.50  & 0.0857 & 0.8383 & 0.0332 \\
        & \rm{GGA}         & 1.2577 & 0.9743 & 1.0319 & 1.8754 & 1.294 & \sim 7.20  & 0.0835 & 0.8115 & 0.0320 \\
        & \rm{LDA+$U$}     & 1.2826 & 0.9897 & 1.0434 & 1.8818 & 1.305 & \sim 9.50  & 0.0754 & 0.8088 & 0.0296 \\
ZnO-z   & \rm{LDA}         & 0.5732 & 0.5552 &        & 0.5883 & 0.577 & \sim 4.60  &        &        &-0.0331 \\
        & \rm{GGA}         & 0.6409 & 0.6152 &        & 0.6494 & 0.638 & \sim 4.60  &        &        &-0.0342 \\
        & \rm{LDA+$U$}     & 1.4859 & 1.4953 &        & 1.4973 & 1.496 & \sim 7.90  &        &        & 0.0020 \\
ZnS-z   & \rm{LDA}         & 1.8745 & 1.8516 &        & 1.9158 & 1.873 & \sim 6.10  &        &        & 0.0642 \\
        & \rm{GGA}         & 2.1134 & 2.0921 &        & 2.1513 & 2.112 & \sim 6.00  &        &        & 0.0592 \\
        & \rm{LDA+$U$}     & 2.3324 & 2.3097 &        & 2.3886 & 2.336 & \sim 9.00  &        &        & 0.0789 \\
&Expt.~[\onlinecite{LB2}]  &        & 3.6800 &        & 3.7400 &       & \geq 9.00  &        &        & 0.0670 \\
&Expt.~[\onlinecite{LB2}]  &        & 3.7800 &        & 3.8500 &       &            &        &        &        \\
ZnSe-z  & \rm{LDA}         & 1.0793 & 0.9484 &        & 1.3409 & 1.079 & \sim 6.60  &        &        & 0.3925 \\
        & \rm{GGA}         & 1.3349 & 1.2089 &        & 1.5858 & 1.335 & \sim 6.50  &        &        & 0.3769 \\
        & \rm{LDA+$U$}     & 1.4214 & 1.2908 &        & 1.6995 & 1.427 & \sim 9.05  &        &        & 0.4087 \\
&Expt.~[\onlinecite{LB2}]  &        & 2.7000 &        &        &       & \sim 9.20  &        &        & 0.4000 \\
&Expt.~[\onlinecite{LB2}]  &        & 2.8200 &        &        &       &            &        &        & 0.4003 \\
ZnTe-z  & \rm{LDA}         & 1.0607 & 0.7715 &        & 1.6681 & 1.070 & \sim 7.10  &        &        & 0.8966 \\
        & \rm{GGA}         & 1.2671 & 0.9862 &        & 1.8533 & 1.275 & \sim 7.05  &        &        & 0.8671 \\
        & \rm{LDA+$U$}     & 1.3287 & 1.0456 &        & 1.9561 & 1.349 & \sim 9.90  &        &        & 0.9105 \\
&Expt.~[\onlinecite{LB2}]  &        & 2.3941 &        &        &       & \sim 9.84  &        &        & 0.9700 \\
&Expt.~[\onlinecite{LB2}]  &        &        &        &        &       & \sim 10.30 &        &        &        \\
\end{tabular}
\end{ruledtabular} \label{band-gap}
\end{table*}
%%%%%%%%%%%%%%%%%%%%%%%%%%%%%%%%%%%%%%%%%%%%%%%%%%%%%%%%%%%%%%%%%%

%%%%%%%%%%%%%%%%%%%%%%%%% Fig%%%%%%%%%%%%%%%%%%%%%%%%%%%%%%%%%%
%%%%%%%%%%%%%%%%%%%%%%%%%%%%%%%%%%%%%%%%%%%%%%%%%%%%%%%%%%%%%%%%%%

\begin{figure}
\vspace{0.0cm}
\includegraphics[scale=3.0]{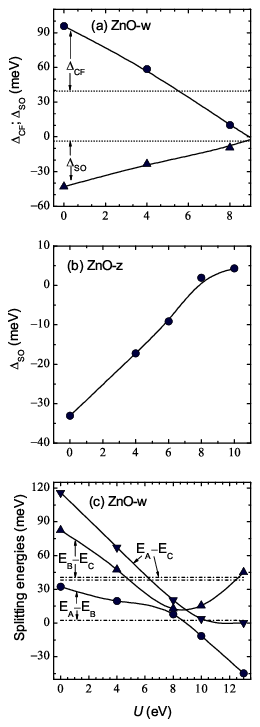}
\setlength{\abovecaptionskip}{0.0cm} \caption{(a) CF splitting, SO
coupling energy for ZnO-w, (b) SO coupling energy for ZnO-z, and
(c) splitting of the states at the top of VB a function of $U$ for
ZnO-w. In panels a and c the solid and dotted lines represent
calculated and experimental data, respectively.}
\label{ZnO+CF+SO+ABC}
\end{figure}

The GGA approach corrected the above mentioned LDA deficiency only
to a little extent. However, LDA+$U$ significantly increased the
values of $E_g, E_A, E_B$, and $E_C$, and shifted the energy
levels of the Zn-3$d$ electrons up to experimentally determined
limits, but only slightly changed the values of $\Delta_{\rm SO}$
and $\Delta_{\rm CF}$ for the Zn$X$ phases except ZnO-z and -w.
The dependence of $\Delta_{\rm CF}$, $\Delta_{\rm SO}$ on $U$ for
ZnO-w is displayed in Fig.~\ref{ZnO+CF+SO+ABC}(a) and of
$\Delta_{\rm SO}$ for ZnO-z in Fig.~\ref{ZnO+CF+SO+ABC}(b).
Analysis of these illustrations shows that $\Delta_{\rm SO}$
remains negative for $U \leq 9.0$~eV for ZnO-w and for $U \leq
8.0$~eV for ZnO-z. For higher values of $U$, $\Delta_{\rm SO}$
becomes a complex number for ZnO-w with a non-zero imaginary part
(in itself physically meaningless), and changes sign for ZnO-z
from negative to positive.

The SO splitting energy is much smaller than 1.0~eV for all phases
except ZnSe-z ($\Delta_{\rm SO} = 0.4$~eV) and ZnTe-z ($0.97$~eV).
Regardless of the computational approach used, the numerical
values of $\Delta_{\rm SO}$ for all Zn$X$ phases remained almost
unchanged. The present values for $\Delta_{\rm SO}$ for the
Zn$X$-z phases are in good agreement with theoretical calculations
[\onlinecite{CW04}] by the LAPW method and available experimental
data.

%%%%%%%%%%%%%%%%%%%%%%%%% Fig%%%%%%%%%%%%%%%%%%%%%%%%%%%%%%%%%%
%%%%%%%%%%%%%%%%%%%%%%%%%%%%%%%%%%%%%%%%%%%%%%%%%%%%%%%%%%%%%%%%%%

The variation of the energy splitting between the $A, B,$ and $C$
states, viz. $E_A-E_B, E_A-E_C,$ and $E_A-E_C,$ at the top of the
VB of ZnO-w is studied as a function of $U$ as shown in
Fig.~\ref{ZnO+CF+SO+ABC}(c). For $U<9.0$~eV the energy splitting
decreases with increasing $U$. At higher values of $U, E_A-E_B$
becomes negative and decreases, $E_B-E_C$ increases, while
$E_A-E_C$ stays more or less constant.

\vspace{-0.4cm}
\subsection{Density of states}
\vspace{-0.4cm}

\begin{figure}
\centering
\includegraphics[scale=0.6]{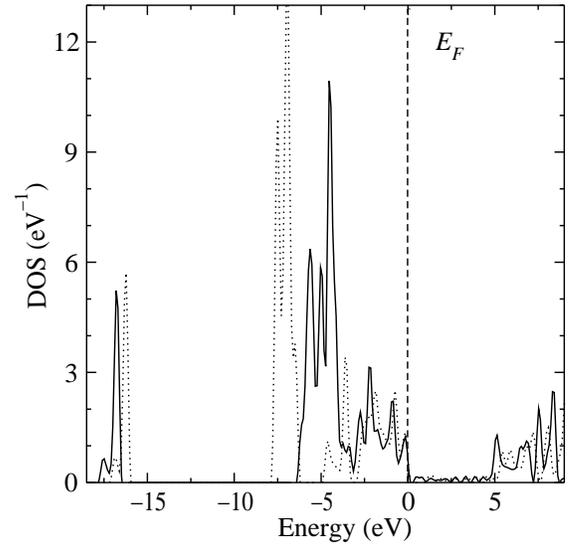}
\caption{Total density of states for ZnO-z calculated from the LDA
(solid line) and LDA+$U$ (dotted line) approaches. Fermi energy
($E_F$) is set at zero of energy.} \label{zZnO+dos}
\end{figure}

The density of states corresponding to the Zn-3$d$~levels
according to the LDA calculations are inappropriately close to the
CB (see e.g. Fig.~\ref{zZnO+dos} for ZnO-z), which contradicts the
findings from XPS, XES, and UPS experiments
[\onlinecite{LPMKS74,VL71,RSS94}]. Furthermore, these states and
the top of the VB are hybridized. Distinct from the other Zn$X$
phases considered, ZnO in both z- and w-type structure shows
artificially widened Zn-3$d$~states. These findings changed only
slightly according to the GGA calculations.

%%%%%%%%%%%%%%%%%%%%%%%%% FIG.3 %%%%%%%%%%%%%%%%%%%%%%%%%%%%%%%%%%
%In old version "scale=0.625"
%%%%%%%%%%%%%%%%%%%%%%%%%%%%%%%%%%%%%%%%%%%%%%%%%%%%%%%%%%%%%%%%%%

The LDA$+U$ approach [\onlinecite{ASKCS93}] was used to adjust the
energy levels of the Zn-3$d$ electrons derived bands to
experimentally established positions using semiempirical values
for the parameters $U$ and $J$. Consequently, the band gaps
calculated by this approach become more reasonable than the pure
LDA-derived band gaps. Further, the height of the peaks in the DOS
corresponding to the Zn-3$d$ states calculated by the LDA+$U$
becomes much larger than those calculated by the pure LDA. This
indicates that the semicore Zn-3$d$ electrons have become more
localized than according to the pure LDA. For ZnO-z and -w the
width of the Zn-3$d$ band calculated by LDA+$U$ becomes much
narrower than that calculated by LDA. For the other Zn$X$ phases,
however, LDA+$U$ only slightly changed the width of the Zn-3$d$
bands, which leads one to conclude that the Coulomb correlation
effects for ZnO is more pronounced than that in other phases
considered.

\vspace{-0.4cm}
\subsection{Order of states at the top of the VB}
\vspace{-0.4cm}

Optical and transport properties for semiconductors strongly
depend on structure of the topmost VB. The order of states at the
top of VB for ZnO-w is one of the topics which is frequently
debated [\onlinecite{RLJLCH99,LRLSM02}]. It is found that the
normal order $\Gamma_5 > \Gamma_1$ is obtained by LDA without SO
coupling for all Zn$X$ phases. Calculations within GGA did not
changed the order. However, the order of the states at the top of
VB for the ZnO-w is changed upon using the LDA+$U$ approach with
the semiempirical values of the parameter $U$, while there appears
no change for the other Zn$X$ phases. For ZnO with SO coupling,
the order of states is $\Gamma_7 > \Gamma_9 > \Gamma_7$, which is
referred to as \textit{anomalous} order, resulting from a negative
SO splitting [\onlinecite{RCP68}].

The variation of the structure at the top of the VB on $U$ with and without SO
coupling is systematically studied for the ZnO-z- and -w-type phases. It is found
that at around the above considered values of $U$ ($U \approx 9.0$~eV for ZnO-w and
$U \approx 8.0$~eV for ZnO-z), the LDA+$U$ interchanges the sequence of the VB
states from $\Gamma_7 > \Gamma_9 > \Gamma_7$ to $\Gamma_7 > \Gamma_7 > \Gamma_9$.

Since in theoretical calculations one can freeze interaction between the valence
electrons and Zn-3$d$ electrons by including the latter into the core, we studied
this particular case for ZnO-z (with SO coupling) and -w (without SO coupling). It
is found that the the Zn-3$d$ electrons are responsible for the order of the states.
On comparing the top of the VB structures calculated within the LDA and GGA
approaches it is found that only small quantitative changes have occurred. Hence,
inhomogeneities in the electron gas do not affect the order of the states at the top
of VB and only slightly changes the band dispersion.

\vspace{-0.4cm}
\subsection{Effective masses}
\vspace{-0.4cm}

The effective masses are calculated along $\Gamma \rightarrow
A(\|)$, and $\Gamma \rightarrow M (\bot)$ with and without the SO
couplings (Table~\ref{em}). According to the conventional
notations, carrier masses for the Zn$X$-z phases are distinguished
by the indices $e$ (electron), $hh$ (heavy-hole), $lh$
(light-hole), and $sh$ (split-off hole). The carrier masses for
the Zn$X$-w phases are distinguished by the indices $e$, $A$, $B$,
and $C$. The calculated CB masses $m_e$ for the Zn$X$-z phases are
more isotropic than those for the Zn$X$-w phases. The numerical
values of $m_e$ for ZnO-w, ZnS-w, ZnSe-z, and ZnTe-z obtained by
the LDA is underestimated by about 50~$\%$ compared to
experimental findings [\onlinecite{H73,Madelung,IOTTK01}], while
those for the other Zn$X$ phases agree fairly well with
experimental data. GGA and LDA+$U$ calculations only slightly
improved the LDA-derived $m_e$ values for all Zn$X$ phases except
ZnO, whereas for ZnO the $m_e$ values calculated from LDA+$U$
gives better agreement with experiment.

The heavy holes along all directions (see Table~\ref{em}) and
light holes along the $\Gamma \rightarrow A(\|)$ direction are
much heavier than other holes and electrons. Hence, the carrier
transport in Zn$X$ is dominated by electrons, while that by holes
can in practice be ruled out. This in turn can be the reason for
the experimentally established large disparity between electron
and hole mobilities [\onlinecite{Madelung}], and also for the
large optical nonlinearity in ZnO [\onlinecite{XC93}]. The hole
effective masses are more anisotropic than those of electrons. On
comparison of the values for $m_e$ in Table~\ref{em} one sees that
the influence of SO coupling on $m_e$ is very important for ZnSe-z
and ZnTe-z, while for the other phases its effect is small.

\begingroup
\squeezetable
\begin{table*}
\caption{Effective masses of electrons and holes (in units of the
free-electron mass $m_0$) for Zn$X$-z and -w phases calculated
within LDA, GGA, and LDA+$U$ approaches. The results are compared
to calculated and experimentally determined data cited in
Ref.~[\onlinecite{Madelung}] and those calculated by FP-LMTO
[\onlinecite{LRLSM02}], LCAO [\onlinecite{XC93}] and determined
experimentally [\onlinecite{H73}]. For ZnO-w labelling of the
effective masses was not changed with changing the order of the
top VB states.}
\begin{ruledtabular}
\begin{tabular}{lldddddddddd}
\multicolumn{1}{l}{} & \multicolumn{1}{l}{} & \multicolumn{1}{c}{$m_e$} & \multicolumn{1}{c}{$m_{hh}^{100}$} & \multicolumn{1}{c}{$m_{hh}^{110}$} & \multicolumn{1}{c}{$m_{hh}^{111}$} & \multicolumn{1}{c}{$m_{lh}^{100}$} & \multicolumn{1}{c}{$m_{lh}^{110}$} & \multicolumn{1}{c}{$m_{lh}^{111}$} & \multicolumn{1}{c}{$m_{SO}^{100}$} & \multicolumn{1}{c}{$m_{SO}^{110}$} & \multicolumn{1}{c}{$m_{SO}^{111}$} \\
\hline \multicolumn{1}{l}{} & \multicolumn{1}{l}{} &
\ \\
ZnO-z & \rm{LDA}            & 0.110  & 0.390  & 0.571  & 0.385  & 1.520 & 1.100  & 1.330  & 0.174  & 0.164 & 0.169 \\
      & \rm{GGA}            & 0.120  & 0.409  & 0.579  & 0.492  & 1.505 & 1.252  & 1.281  & 0.188  & 0.186 & 0.181 \\
      & \rm{LDA}+$U$        & 0.193  & 1.782  & 2.920  & 1.972  & 0.968 & 1.392  & 1.669  & 0.250  & 0.240 & 0.230 \\
ZnS-z & \rm{LDA}            & 0.150  & 0.775  & 1.766  & 2.755  & 0.224 & 0.188  & 0.188  & 0.385  & 0.355 & 0.365 \\
      & \rm{GGA}            & 0.172  & 0.783  & 1.251  & 3.143  & 0.233 & 0.216  & 0.202  & 0.378  & 0.373 & 0.383 \\
      & \rm{LDA}+$U$        & 0.176  & 1.023  & 1.227  & 1.687  & 0.268 & 0.252  & 0.218  & 0.512  & 0.445 & 0.447 \\
&Expt.~[\onlinecite{Madelung}]& 0.184  &        &        & 1.760  &       & 0.230  &        &        &       &       \\
&Expt.~[\onlinecite{Madelung}]& 0.340  &        &        &        &       &        &        &        &       &       \\
ZnSe-z& \rm{LDA}            & 0.077  & 0.564  & 1.310  & 1.924  & 0.104 & 0.100  & 0.094  & 0.250  & 0.246 & 0.254 \\
      & \rm{GGA}            & 0.098  & 0.568  & 0.922  & 1.901  & 0.126 & 0.122  & 0.111  & 0.271  & 0.273 & 0.267 \\
      & \rm{LDA}+$U$        & 0.100  & 0.636  & 1.670  & 1.920  & 0.129 & 0.120  & 0.117  & 0.287  & 0.297 & 0.309 \\
&Expt.~[\onlinecite{Madelung}]& 0.130  & 0.570  & 0.750  &        &       &        &        &        &       &       \\
&Expt.~[\onlinecite{Madelung}]& 0.170  &        &        &        &       &        &        &        &       &       \\
ZnTe-z& \rm{LDA}            & 0.064  & 0.381  & 0.822  & 1.119  & 0.071 & 0.067  & 0.066  & 0.254  & 0.253 & 0.256 \\
      & \rm{GGA}            & 0.078  & 0.418  & 0.638  & 1.194  & 0.093 & 0.086  & 0.081  & 0.261  & 0.255 & 0.274 \\
      & \rm{LDA}+$U$        & 0.081  & 0.483  & 0.929  & 1.318  & 0.096 & 0.088  & 0.085  & 0.288  & 0.292 & 0.290 \\
&Expt.~[\onlinecite{Madelung}]& 0.130  &        & 0.600  &        &       &        &        &        &       &       \\
\ \\
\hline
\multicolumn{1}{l}{} & \multicolumn{1}{l}{} & \multicolumn{1}{c}{$m_e^{\|}$} & \multicolumn{1}{c}{$m_e^{\bot}$} & \multicolumn{1}{c}{$m_A^{\|}$} & \multicolumn{1}{c}{$m_A^{\bot}$} & \multicolumn{1}{c}{$m_B^{\|}$} & \multicolumn{1}{c}{$m_B^{\bot}$} & \multicolumn{1}{c}{$m_C^{\|}$} & \multicolumn{1}{c}{$m_C^{\bot}$} & \multicolumn{1}{l}{} & \multicolumn{1}{l}{} \\
\hline
ZnO-w & \rm{LDA}               & 0.137 & 0.130 & 2.447 & 2.063 & 2.979 & 0.227 & 0.169 & 0.288 &&\\
      & \rm{GGA}               & 0.144 & 0.143 & 2.266 & 0.351 & 3.227 & 0.300 & 0.165 & 0.537 &&\\
      & \rm{LDA}+$U$           & 0.189 & 0.209 & 0.207 &11.401 & 4.330 & 3.111 & 0.330 & 0.270 &&\\
&FP-LMTO.~[\onlinecite{LRLSM02}]& 0.230 & 0.210 & 2.740 & 0.540 & 3.030 & 0.550 & 0.270 & 1.120 &&\\
&Expt.~[\onlinecite{H73}]      & 0.24  &       & 0.590 & 0.590 & 0.590 & 0.590 & 0.310 & 0.550 &&\\
&LCAO.~[\onlinecite{XC93}]     & 0.280 & 0.320 & 1.980 & 4.310 &       &       &       &       &&\\
ZnS-w & \rm{LDA}               & 0.144 & 0.153 & 1.746 & 3.838 & 0.756 & 0.180 & 0.183 & 0.337 &&\\
      & \rm{GGA}               & 0.142 & 0.199 & 2.176 & 1.713 & 0.402 & 0.198 & 0.440 & 0.443 &&\\
      & \rm{LDA}+$U$           & 0.138 & 0.157 & 1.785 & 2.194 & 0.621 & 0.195 & 0.339 & 0.303 &&\\
&Expt.~[\onlinecite{Madelung}] & 0.280 &       & 1.400 & 0.490 &       &       &       &       &&\\
&LCAO.~[\onlinecite{XC93}]     & 0.260 & 0.330 & 1.510 & 1.470 &       &       &       &       &&\\
ZnSe-w& \rm{LDA}               & 0.148 & 0.139 & 1.404 & 0.158 & 0.114 & 0.124 & 0.171 & 0.197 &&\\
      & \rm{GGA}               & 0.184 & 0.149 & 1.395 & 0.184 & 0.135 & 0.173 & 0.190 & 0.306 &&\\
      & \rm{LDA}+$U$           & 0.185 & 0.149 & 1.629 & 0.189 & 0.137 & 0.187 &       & 0.344 &&\\
ZnTe-w& \rm{LDA}               & 0.108 & 0.128 & 1.042 & 0.118 & 0.070 & 0.105 & 0.229 & 0.237 &&\\
      & \rm{GGA}               & 0.134 & 0.182 & 1.044 & 0.122 & 0.102 & 0.145 & 0.239 & 0.246 &&\\
      & \rm{LDA}+$U$           & 0.131 & 0.184 & 1.116 & 0.131 & 0.128 & 0.166 &       &       &&\\
&Ref.~[\onlinecite{Madelung}]  & 0.130 &       & 0.600 &       &       &       &       &       &&\\
\end{tabular}
\end{ruledtabular}  \label{em}
\end{table*}
\endgroup

\vspace{-0.4cm}
\section{Conclusion}
\vspace{-0.4cm}

Electronic structure and band parameters for Zn$X$-w and -z phases
are studied by first-principles calculations within the LDA, GGA,
and LDA+$U$ approaches. It is found that LDA underestimates the
band gaps, the actual positions of the energy levels of the
Zn-3$d$ states  as well as splitting energies between the states
at the top of VB, but overestimates the crystal-field splitting
energy. The spin-orbit coupling energy is overestimated for ZnO-w,
underestimated for ZnS-w, and comes more or less accurate for
ZnS-z, ZnSe-z, and ZnTe-z.

The LDA+$U$ calculation as a function of $U$ and $J$ has been used
to adjust the Zn-3$d$ band derived from LDA to the experimentally
established location of the Zn-3$d$ levels determined from X-ray
emission spectra. Using the $U$ values corresponding to the
experimentally found $E_d$ location the calculated band gaps and
band parameters are improved compared to pure LDA approach.

The order of the states at the top of VB is systematically
examined for ZnO-z and -w phases. It is shown that without SO
coupling the band order is normal, but it becomes anomalous and
$\Delta_{\rm{SO}}$ goes negative upon inclusion of SO coupling. It
is found that in the LDA+$U$ calculations the anomalous order is
maintained until $U \approx\ 8.0$~eV for ZnO-z and $U \approx\
9.0$~eV for ZnO-w. Above these $U$  values, the band order is
inverted and becomes normal and $\Delta_{\rm{CF}}$ for ZnO-w goes
from positive to negative, whereas $\Delta_{\rm{SO}}$ converts to
a complex quantity. It indicates that either the quasicubic model,
within which $\Delta_{\rm{CF}}$ and $\Delta_{\rm{SO}}$ are
calculated, do not work for this particular case or such order is
unphysical.

Upon excluding the interaction between the Zn-3$d$ and other
valence electrons by including the former in the core, the order
becomes anomalous. Based on these analyses one can conclude that
the Zn-3$d$ electrons are responsible for the anomalous order of
the states at the top of VB in ZnO.

Effective masses of electrons at the conduction-band minimum and
of holes at the valence-band maximum have been calculated. The
heavy holes in the VB are found to be much heavier than the CB
electrons. The calculations, moreover, indicate that effective
masses of the holes are much more anisotropic than those of the
electrons. CB electron masses for ZnO-w, ZnS-w, ZnSe-z, and ZnTe-z
calculated within LDA are underestimated by about 50~$\%$ compared
with experimental data, while those for the other Zn$X$ phases
considered agree with experimental data.

The GGA approach does not remedy the LDA derived error in energy
gaps and band parameters. We found that SO coupling is important
for calculation of the parameters for the z-ZnSe and z-ZnTe
phases, while it is not significant for other Zn$X$ phases.
However, the calculated $\Delta_{\rm CF}$ values within the
different approaches do not differ much except for ZnO emphasizing
that Coulomb correlation effects are more pronounced for ZnO than
ZnS, ZnSe, and ZnTe.

\vspace{-0.4cm}
\section*{Acknowledgments}
\vspace{-0.4cm}

This work has received financial and supercomputing support from
the Research Council of Norway. SZK thanks R.~Vidya, P.~Vajeeston
and A.~Klaveness (Department of Chemistry, University of Oslo) for
discussions and assistance. We also thank Professor~M.A.~Korotin
(Institute of Metal Physics, Ekaterinburg, Russia) for help with
the computations of the values of the parameters U and J within
the constrain DFT.

%\bibliographystyle{apsrev}
%\bibliography{C:/Smagul/Papers/Prb1/PRB1/Apl}

\end{document}